\documentclass[a4paper,11pt]{article}
\pdfoutput=1

\usepackage{jinstpub}

\usepackage{lineno}
\usepackage{siunitx}
\usepackage{booktabs}
\usepackage{xcolor}
\usepackage{float}
\usepackage{subcaption}
\RequirePackage[numbers,sort&compress]{natbib}

\title{\boldmath Machine Learning Based ROI Segmentation for Beam Imaging Diagnostics at Accelerators}

\author[a]{Prachiti Sujit Chandratreya}
\emailAdd{chandratreyaprachiti1618@gmail.com}
\author[b]{Frank Mayet}
\emailAdd{frank.mayet@desy.de}
\author[b]{Sergey Tomin}
\emailAdd{sergey.tomin@desy.de}
\author[a,1]{Jitendra Kumar,\note{Corresponding author.}}
\emailAdd{jkumar@iitj.ac.in}

\affiliation[a]{Indian Institute of Technology Jodhpur,\\
Karwar, Jodhpur, Rajasthan, India}
\affiliation[b]{Deutsches Elektronen-Synchrotron (DESY),\\
Notkestrase 85, 22607 Hamburg, Germany}

\abstract{The European XFEL accelerator produces high-brightness X-ray pulses using relativistic electron beams. Beam characterization is performed at multiple diagnostic stations using scintillator screens, where beam images are analyzed to extract key parameters such as emittance, energy spread, and current profile. Accurate detection of the region of interest in these images is essential for reliable beam diagnostics and stable accelerator operation. Conventional methods, such as bounding-box-based beam localization, can become less reliable for complex or non-ideal beam profiles, such as low-intensity signals, tilted or streaked beams, and multiple beam structures. In this work, we develop machine learning-based approaches for region of interest detection directly from beam images. These ML methods adapt to variations in beam shape and intensity and enable more precise, pixel-level identification of beam regions. Experimental results demonstrate improved robustness and accuracy in challenging conditions compared to traditional techniques. The proposed methodology is validated at the European XFEL and is broadly applicable to image-based beam diagnostics across accelerator facilities.}

\keywords{Accelerator applications, Beam position and profile monitors, Image segmentation, Pattern recognition, Machine learning}
\begin{document}
\maketitle
\flushbottom

\section{Introduction}
\label{sec:intro}
X-ray Free Electron Lasers (XFELs) are advanced accelerator-based light sources that generate ultra-bright, femtosecond X-ray pulses for probing matter at atomic and molecular scales~\cite{emma2010first, decking2020mhz}. The European XFEL operates using high-energy electron bunches accelerated up to 17.5~GeV in a superconducting linear accelerator and subsequently passed through long undulator sections,  where coherent X-ray radiation is generated via the self-amplified spontaneous emission (SASE) process~\cite{bonifacio1984sase, xfel_tdr, tschentscher2017photon}. This work focuses on the European XFEL; however, the beam image analysis challenges addressed here are common across a wide range of accelerator facilities. Consequently, the proposed methodology is broadly applicable to image-based beam diagnostics in particle accelerators.

The generation of precisely controlled electron beams is central to XFEL operation. These high-energy, relativistic electron bunches ultimately dictate the properties of the generated X-ray radiation, requiring well-defined spatial and temporal characteristics. Accurate characterization of the electron beam is essential for stable operation and optimal performance of XFEL facilities. At the European XFEL, beam diagnostics are primarily performed using scintillator screens installed at multiple diagnostic stations, where recorded images are analyzed to extract key beam properties such as beam size and intensity distribution~\cite{walasek2012scintillating, wiebers2013scintillating}, and to support advanced beam diagnostic techniques including energy spread and longitudinal phase-space measurements~\cite{tomin2021accurate, tomin2022longitudinal, Dijkstal2024}.

A critical step in beam image analysis is the online identification of the region of interest (ROI), corresponding to the portion of the image containing the beam signal. Reliable ROI detection is essential to suppress background noise and enable accurate extraction of beam properties such as intensity and beam spread. However, beam images frequently exhibit complex structures, including low-intensity signals, distorted or tilted profiles, streaking effects, and multiple bunches, which limit the performance of conventional image-processing techniques.

To address these limitations, robust and reliable ROI detection methods are required. Conventional beam localization approaches, such as bounding-box methods, are computationally efficient but assume relatively simple beam geometries and often fail under realistic operating conditions. In this work, we investigate machine-learning-based approaches for robust, pixel-level ROI detection directly from beam images. These methods adapt to variations in beam shape and intensity, enabling more accurate and reliable ROI identification across a wide range of operating conditions.

\section{Beam Image Analysis and ROI Detection}
In beam diagnostics at XFEL facilities, the region of interest (ROI) corresponds to the portion of the image containing the electron beam signal, while excluding background noise and artifacts. Accurate identification of the ROI is essential for reliable beam image analysis and downstream extraction of beam properties. At the European XFEL, ROI detection is currently performed within the DESY Image Analysis framework using bounding-box methods. These approaches identify the beam by enclosing regions of high intensity within rectangular boundaries. While computationally efficient and suitable for real-time operation, they rely on assumptions of compact and well-defined beam shapes.

In practice, beam images often deviate significantly from ideal conditions. Typical challenges include low-intensity or low charge-density beams, distorted or tilted beam profiles, streaking effects arising from diagnostic elements such as transverse deflecting structures, and the presence of multiple beam structures (e.g., twin-bunch configurations), where overlapping regions complicate reliable ROI detection. Under such conditions, conventional methods may fail to capture the full beam extent or may incorrectly include background regions. Classical image-processing techniques, such as threshold-based image segmentation~\cite{otsu1979threshold, gonzalez2018digital}, were also investigated in this work. These methods can adapt to intensity variations and detect finer beam structures under controlled conditions. However, when evaluated on both synthetic and real XFEL beam images, segmentation approaches showed limited robustness. While they perform well on synthetic data with well-defined beam features, they tend to overestimate the ROI in real experimental images due to sensitivity to noise and background fluctuations.

\section{ROI Detection Methods}

\subsection{Synthetic Dataset Generation}
During operation, the European XFEL produces a large volume of beam image data. However, these data do not contain ground truth ROI masks, which are required for supervised learning. Therefore, a synthetic dataset was developed to enable controlled training and evaluation of machine learning models. The synthetic data generation pipeline produces realistic beam images with controlled variability in beam shape, intensity, position, overlap, and noise characteristics. This enables systematic evaluation of model performance under diverse beam conditions, including challenging scenarios such as faint beams, distorted profiles, and multiple beam structures.

Synthetic beam images were generated by starting from a two-dimensional array with zero intensity and adding beam features, such as streaks and blobs, using parameterized two-dimensional Gaussian distributions. For each of these images, augmented versions were also generated, containing artificial noise to closely mimic accelerator conditions, with the same ground truth mask. These images include various types of signals and background/noise. Background noise represents the primary challenge in real beam images, so we made it a more realistic background with Gaussian noise, gradients, readout patterns, and hot pixels. The ROIs were modeled as streaks, spots, arcs, diffuse blobs, and rings, with parameters such as size, rotation, intensity, and position chosen randomly. To simulate faint ROIs, the intensity was kept low (0.1--0.6), producing subtle, low-contrast regions. Augmentations such as flips, rotations, brightness and contrast changes, elastic distortions, and grid distortions were also applied.

\begin{figure}[h]
    \centering
    \includegraphics[width=0.80\textwidth]{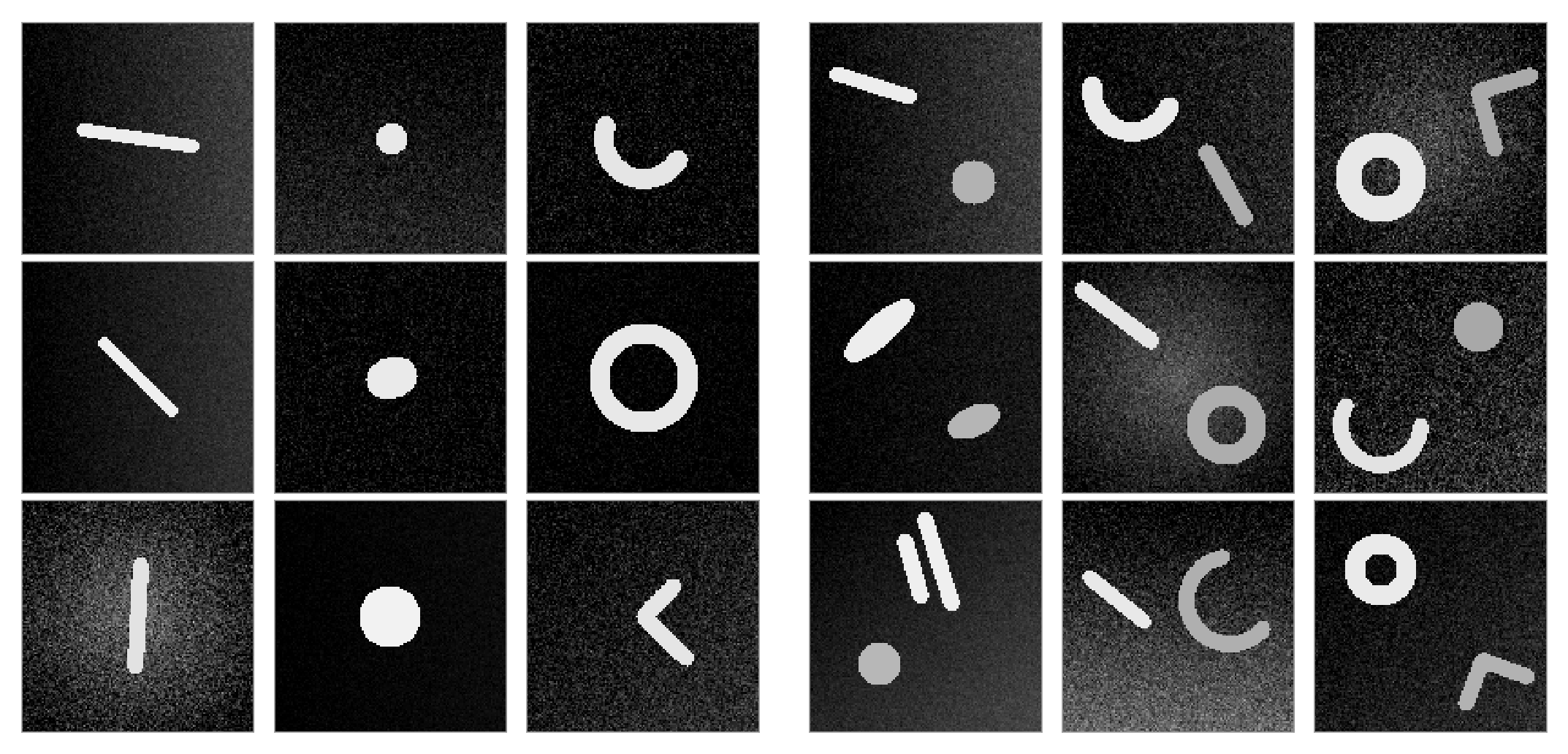}
    \caption{Examples of synthetically generated images used for machine learning training, illustrating diverse beam profiles and noise conditions.}
    \label{fig:data1}
\end{figure}

The dataset consists of several thousand images with corresponding ground truth ROI masks. For the autoencoder and multiclass U-Net models, the dataset was divided into training and validation sets using an 80--20 split. For the single-class U-Net, a relatively larger dataset was used, divided into training, validation, and test sets using a 70-15-15 split, with the held-out test set used to obtain an unbiased estimate of final model performance. Since the training data were generated synthetically, a domain gap may exist between the synthetic dataset and real XFEL beam images. In particular, experimental images may contain previously unseen noise characteristics, detector artifacts, or beam features that are not fully represented in the synthetic data. To mitigate this effect, the synthetic images were designed to approximate the intensity distribution and noise characteristics observed in XFEL beam images. The trained models were subsequently evaluated on real XFEL beam images to assess their generalization capability. Fig.~\ref{fig:data1} shows examples of synthetically generated beam images used for model training.

\subsection{Machine Learning Models}
Three machine learning approaches were investigated for ROI detection: single-class U-Net, multi-class U-Net, and autoencoder-based models. All models were initially developed and trained on the DESY Maxwell GPU cluster using NVIDIA A100 GPUs; however, the results reported in this work were obtained by training and testing all three models on cloud GPU resources (NVIDIA Tesla T4, via Kaggle).

\subsubsection{Single-Class U-Net}
U-Net is a convolutional neural network designed for image segmentation, using an encoder--decoder architecture with skip connections that enable precise pixel-level ROI detection~\cite{ronneberger2015unet,minaee2022survey}. This makes it well-suited for pixel-level ROI detection in beam images. The encoder progressively extracts hierarchical features through convolutional blocks with batch normalization, while the decoder reconstructs the segmentation mask using upsampling and skip connections from corresponding encoder layers. This structure allows the model to capture both global context and fine spatial details, which is essential for detecting faint and complex beam structures.

The implemented U-Net model uses encoder feature sizes up to 512 channels and a bottleneck of 1024 channels to enhance the representation of subtle beam features. Batch normalization is applied at each convolutional stage to stabilize training. 
The model is trained using a combination of Binary Cross-Entropy (BCE) and Dice loss. Dice loss helps mitigate class imbalance by directly optimizing the overlap between predicted and ground-truth regions~\cite{milletari2016vnet}. BCE optimizes pixel-wise classification, while Dice loss emphasizes overlap between the predicted and ground-truth regions, encouraging more accurate segmentation of small foreground regions. Data augmentation techniques, including rotation, flipping, elastic deformation, and intensity scaling, are applied to improve robustness to variations in beam shape and noise. The overall architecture and training configuration are summarized in Table~\ref{tab:unet_comparison}.

\begin{table}[H]
\centering
\caption{U-Net architecture and training configuration.}
\begin{tabular}{|p{3cm}|p{5.5cm}|p{5cm}|}
\hline
\textbf{Aspect} & \textbf{U-Net (Optimized)} & \textbf{Effect} \\
\hline
Encoder filters & 512 (DoubleConv + BatchNorm) & Rich feature extraction \\
\hline
Bottleneck & 1024 channels & Strong latent representation \\
\hline
Regularization & BatchNorm, no dropout & Stable training \\
\hline
Decoder filters & 64 with skip connections & Sharp reconstruction \\
\hline
Dataset & Synthetic + augmentation & Robustness to variation \\
\hline
Loss & BCE + Dice & Improved structural accuracy \\
\hline
\end{tabular}
\label{tab:unet_comparison}
\end{table}

This combination of architecture and training strategy enables the U-Net model to achieve accurate and robust ROI segmentation across a wide range of beam conditions. The U-Net architecture is particularly effective for this task due to its ability to preserve spatial information through skip connections while capturing global beam structure, enabling reliable detection of faint and distorted ROI features.

\begin{figure}[H]
    \centering
\includegraphics[width=0.97\textwidth]{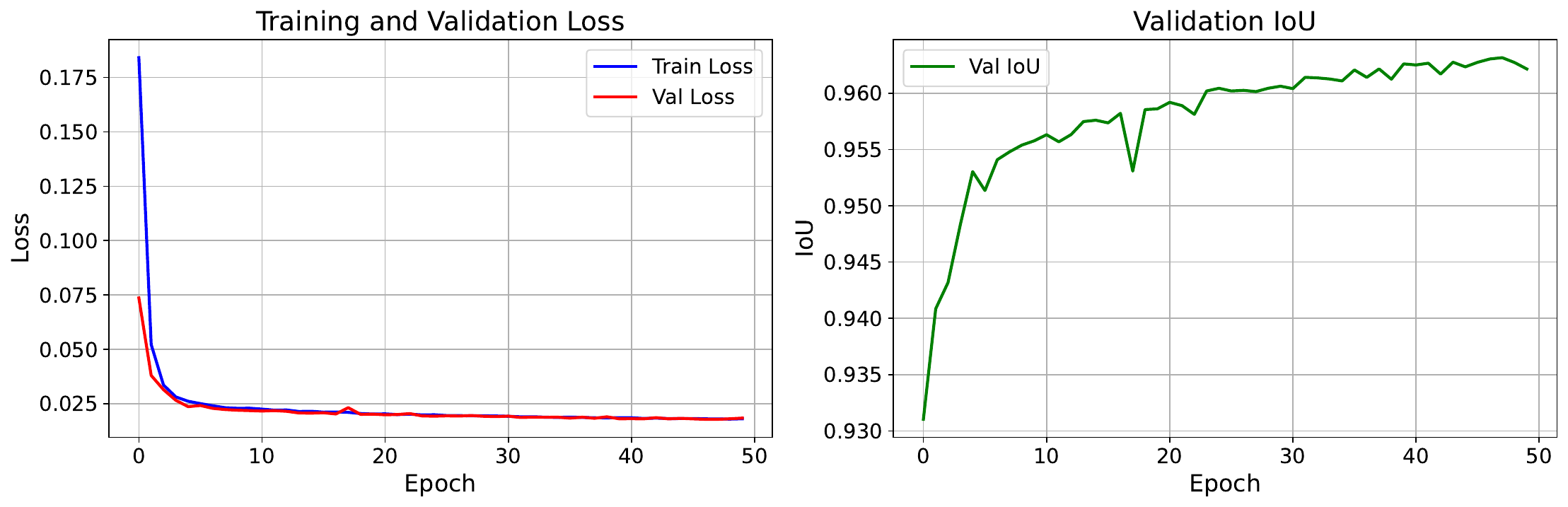}
    \caption{Training and validation loss curves and validation IoU of the Single-Class U-Net model over 50 epochs, showing stable convergence and strong segmentation performance.}
    \label{fig:unet_training}
\end{figure}

\begin{figure}[H]
    \centering   
    \includegraphics[width=0.55\textwidth]{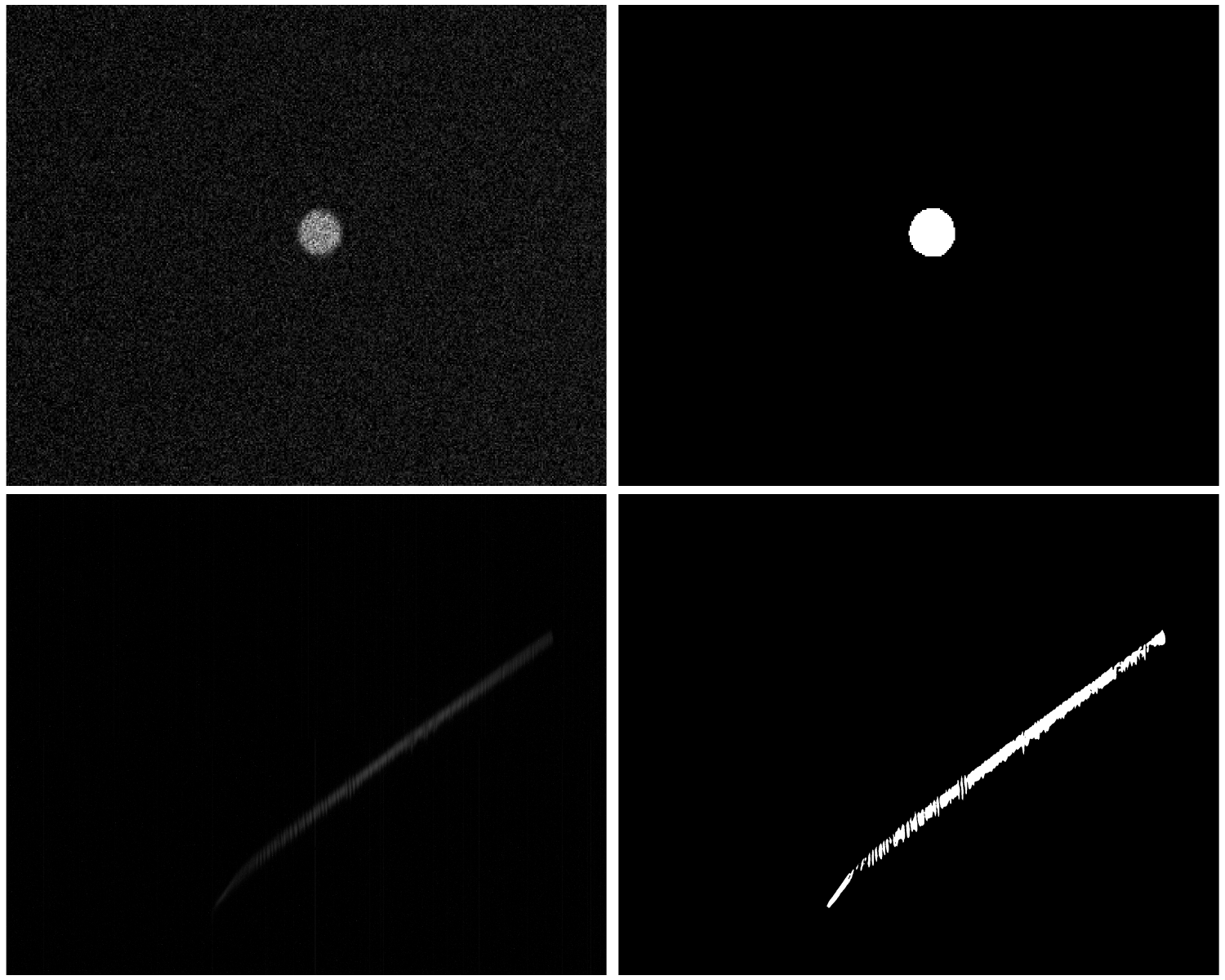}
    \caption{Single-Class U-Net ROI detection results for synthetic beam images (top) and real XFEL accelerator beam images (bottom). In each pair, the left image shows the input beam image and the right image shows the corresponding predicted ROI mask.}    
    \label{fig:unet_roi_comparison}
\end{figure}

The single-class U-Net model achieved stable convergence during training, as shown in Fig.~\ref{fig:unet_training}, with consistently decreasing training and validation losses and increasing validation Intersection over Union (IoU), which measures the overlap between the predicted segmentation and the ground-truth mask. Representative ROI detection results for both synthetic and real XFEL accelerator beam images are presented in Fig.~\ref{fig:unet_roi_comparison}, demonstrating accurate segmentation performance under different beam conditions. The quantitative evaluation results summarized in Table~\ref{tab:single_unet_results} further confirm the effectiveness of the model, achieving an IoU of 0.9625 and an accuracy score of 0.9946.

\begin{table}[ht]
\centering
\caption{Segmentation performance metrics of the single-class U-Net model for ROI detection on the held-out test set.}
\label{tab:single_unet_results}
\begin{tabular}{lcccc}
\hline
Model & IoU $\uparrow$ & Accuracy $\uparrow$ & Precision $\uparrow$ & Recall $\uparrow$ \\
\hline
U-Net (Single-class) & 0.9625 & 0.9946 & 0.9825 & 0.9793 \\
\hline
\end{tabular}
\end{table}

\subsubsection{Autoencoder}
Autoencoders are neural networks that learn compact latent representations of input data for feature extraction and representation learning~\cite{hinton2006reducing}. Denoising autoencoders extend this framework by learning to reconstruct clean data from corrupted inputs, thereby enabling effective noise suppression~\cite{vincent2010stacked}. In this work, a denoising autoencoder is trained to reconstruct clean beam images from noisy inputs, enhancing the beam signal and increasing ROI visibility. An enhanced architecture incorporating a deep bottleneck structure and mixed-precision training was also implemented to improve computational efficiency and stability. The enhanced architecture demonstrated stable convergence and effective reconstruction capability for beam profile analysis.

\begin{figure}[H]
    \centering
    \includegraphics[width=0.75\textwidth]{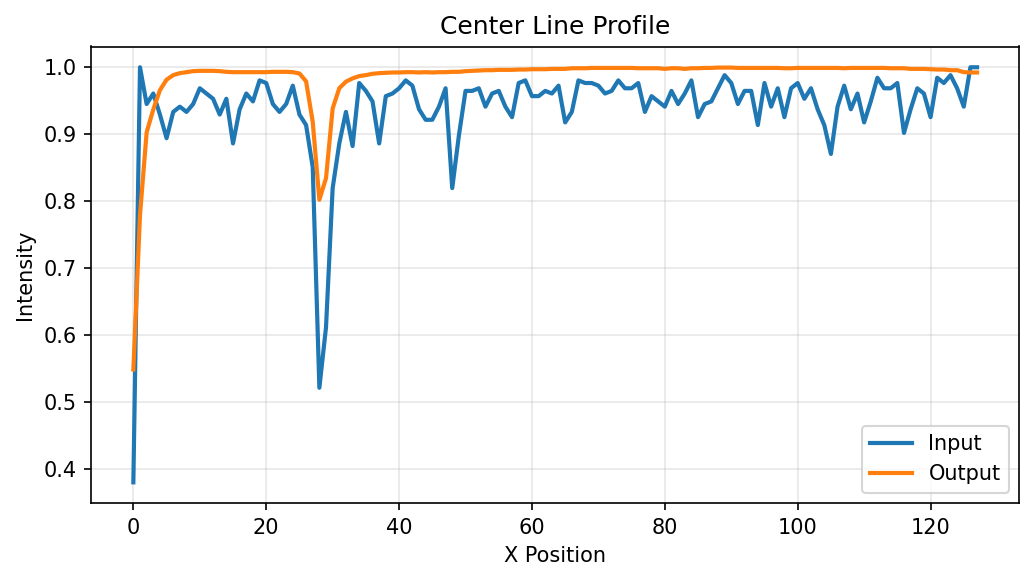}
    \caption{Center-line intensity profile of a twin-bunch XFEL beam reconstructed by the enhanced autoencoder. The original noisy signal (blue) and the autoencoder output (orange) are shown; the model retains the principal beam structure while attenuating local noise fluctuations.}
    \label{fig:autoencoder_training}
\end{figure}

\begin{figure}[H]
    \centering
    
    \begin{minipage}{0.35\textwidth}
        \centering
        \includegraphics[width=\textwidth]{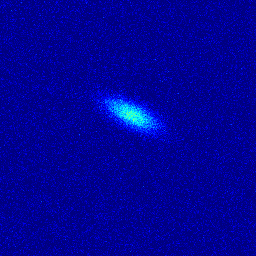}
        % \caption*{(a) Synthetic input image}
    \end{minipage}
    \begin{minipage}{0.35\textwidth}
        \centering
        \includegraphics[width=\textwidth]{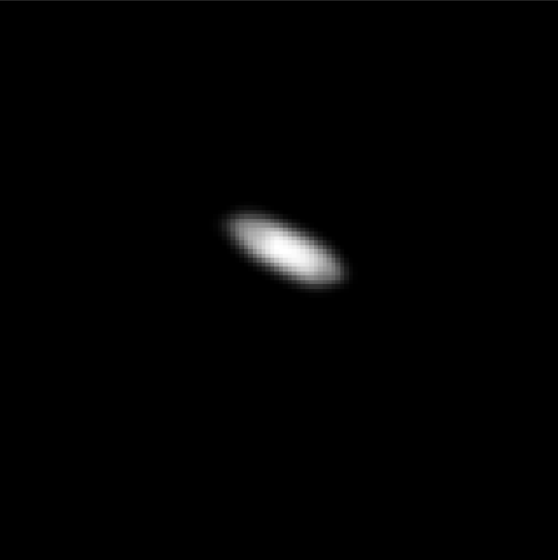}
        % \caption*{(b) Output mask generated}
    \end{minipage}
    \caption{Example of a synthetic input image (left) and the corresponding segmentation mask (right) predicted by the enhanced autoencoder.}
    \label{fig:autoencoder_output}
\end{figure}

Fig.~\ref{fig:autoencoder_training} shows the reconstructed center-line intensity profile, where the autoencoder output closely follows the input signal while reducing local fluctuations and noise artifacts. These results indicate that the model preserves the main beam structure while improving signal smoothness and robustness. The autoencoder model processes grayscale beam images and generates reconstructed outputs or probability maps that are thresholded to obtain ROI masks. While the autoencoder effectively suppresses noise and enhances beam structures, its segmentation accuracy is lower than that of the U\text{-}Net model because it is optimized primarily for image reconstruction rather than direct segmentation. Nevertheless, the method provides useful preprocessing and feature enhancement for complex XFEL beam images. The network operates on \(128 \times 128\) grayscale inputs using an encoder--decoder architecture with a latent bottleneck representation and sigmoid activation applied to network outputs. Binary ROI masks are generated by thresholding the output probability map at 0.5. An example synthetic input image and the corresponding predicted ROI mask are shown in Fig.~\ref{fig:autoencoder_output}. 

\begin{figure}[H]
\centering
\begin{minipage}{0.31\textwidth}
    \centering
    \includegraphics[width=\linewidth,height=4.2cm]{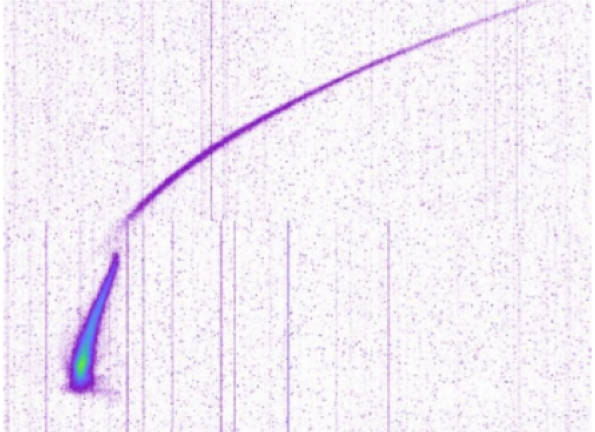}
\end{minipage}
\hfill
\begin{minipage}{0.31\textwidth}
    \centering
    \includegraphics[width=\linewidth,height=4.2cm]{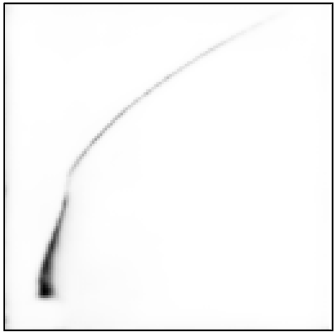}
\end{minipage}
\hfill
\begin{minipage}{0.31\textwidth}
    \centering
    \includegraphics[width=\linewidth,height=4.2cm]{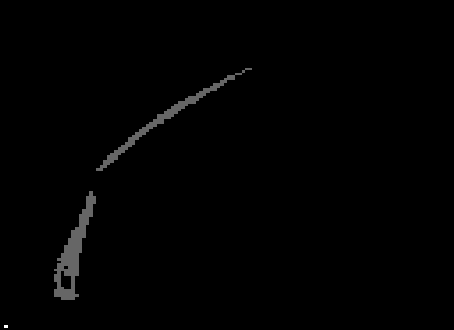}
\end{minipage}
\caption{Comparison of autoencoder- and multiclass U-Net-based ROI analysis for a real twin-bunch XFEL beam image. Left: input beam image. Center: reconstructed beam probability map generated by the enhanced autoencoder. Right: multiclass ROI mask predicted by the U-Net, showing the separated beam structures.}
\label{fig:twinbunch_model_comparison}
\end{figure}

\begin{table}[H]
\centering
\caption{Autoencoder architecture and training configuration.}
\begin{tabular}{|p{3cm}|p{5.5cm}|p{5cm}|}
\hline
\textbf{Aspect} & \textbf{Autoencoder (Enhanced)} & \textbf{Effect} \\
\hline
Input size & 128 $\times$ 128 grayscale & Standardized processing \\
\hline
Encoder depth & 4 convolutional blocks & Progressive feature extraction \\
\hline
Bottleneck & 8 $\times$ 8 $\times$ 512, 256:1 spatial reduction & Strong latent representation, noise suppression \\
\hline
Skip connections & Additive skip connections & Preserves fine spatial details \\
\hline
Decoder & Symmetric upsampling layers & Clean ROI reconstruction\\
\hline
Loss function & Weighted BCE + structural losses & Balances accuracy and structural preservation \\
\hline
Training precision & Mixed precision (AMP) & Faster training, reduced memory usage \\
\hline
Augmentation & Flips, rotations, brightness jitter, noise injection & Improved robustness to real beam artifacts \\
\hline
\end{tabular}
\label{tab:autoencoder_comparison}
\end{table}

The framework additionally provides input--output comparisons, confidence heatmaps, noise suppression visualizations, and quantitative beam metrics including pixel count, coverage percentage, and intensity distribution. Representative reconstruction results of the enhanced autoencoder for a real twin-bunch XFEL beam image are shown in the center panel of Fig.~\ref{fig:twinbunch_model_comparison}, while the overall architecture and training configuration are summarized in Table~\ref{tab:autoencoder_comparison}.

\subsubsection{Multiclass U-Net}

To address complex beam structures such as twin bunches, two approaches were investigated: intensity-based post-processing and multiclass U-Net segmentation. In the multiclass framework, each pixel was assigned to the background or to different ROI categories, enabling direct separation of multiple beam regions within a single image. Compared to threshold-based post-processing, the multiclass U-Net provided more consistent separation of overlapping beam structures and improved robustness for complex intensity distributions. The proposed models were evaluated using synthetic beam images containing varying beam shapes, noise levels, intensity distributions, and overlapping structures. Under these conditions, the multiclass U-Net achieved accurate ROI detection across diverse beam profiles, including elongated and low-intensity beams, while effectively suppressing background noise. Fig.~\ref{fig:twinbunch_model_comparison} compares the enhanced autoencoder and the multiclass U-Net on a real twin-bunch XFEL beam image, illustrating the improved separation of overlapping beam structures achieved by the multiclass approach.

\begin{table}[H]
\centering
\caption{Performance comparison of enhanced autoencoder and multiclass U-Net models.}
\begin{tabular}{lccc p{3.5cm}}
\hline
Model & Train Loss $\downarrow$ & Val Loss $\downarrow$ & IoU $\uparrow$ & Remarks \\
\hline
Enhanced Autoencoder 
& 0.1313 
& 0.1319 
& 0.6536
& No explicit class-wise segmentation \\
\hline
Multiclass U-Net 
& 0.0280 
& 0.0004 
& \begin{tabular}[c]{@{}c@{}}
Faint: 0.9996 \\
Bright: 0.9998
\end{tabular}
& Class-wise ROI detection \\
\hline
\end{tabular}
\label{tab:autoencoder_multiclass}
\end{table}

Table~\ref{tab:autoencoder_multiclass} compares the enhanced autoencoder and multiclass U-Net models. The enhanced autoencoder produced smooth denoised reconstructions with stable training and validation losses; however, its reconstruction-oriented design resulted in blurred ROI boundaries and limited separation of overlapping beam regions, making it less suitable for complex beam structures requiring class-wise interpretation. In contrast, the multiclass U-Net achieved improved boundary localization and accurate separation of faint and bright beam regions through explicit semantic segmentation, resulting in high IoU values for both ROI classes while preserving structural and intensity-related information. Although the single-class U-Net improved ROI localization compared with the autoencoder, its binary segmentation formulation could not distinguish different beam intensity regions. These results demonstrate that, while autoencoder-based methods are effective for denoising and coarse ROI reconstruction, the multiclass U-Net provides a more suitable framework for accurate XFEL beam diagnostics and analysis of complex beam structures.

\section{Performance on XFEL server}
In addition to accuracy, computational performance is critical for practical deployment. The optimized U-Net model was deployed within the DESY computing environment and evaluated on both GPU and CPU platforms. On the Maxwell GPU cluster, the model achieves inference speeds suitable for real-time applications. Further evaluation on a control system development platform using an Apple Mac mini (M4 Pro) showed inference rates exceeding 10 Hz using Metal Performance Shaders (MPS), while CPU-based inference required approximately 0.5 second per frame. The computational performance on different hardware platforms is summarized in Table~\ref{tab:computational_performance}.

\begin{figure}[H]
    \centering
    \includegraphics[width=0.7\textwidth]{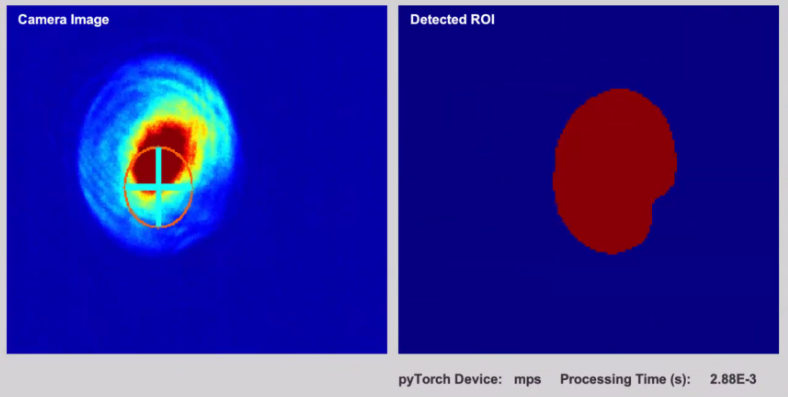}
    \caption{Real-time ROI detection system deployed on the server, showing live beam images and corresponding predicted ROI masks.}
    \label{fig:server_inference}
\end{figure}

The inference pipeline has been integrated into a prototype control system server that continuously acquires beam images and performs real-time ROI detection. A prototype real-time deployment of the ROI detection system with the U-Net model running on the server is shown in Fig.~\ref{fig:server_inference}. The system operates in synchronization with the machine timing system and has been successfully tested under beam conditions. These results demonstrate that the proposed approach is not only accurate but also sufficiently efficient for online accelerator diagnostics and control applications.

\begin{table}[h!]
\centering
\caption{Computational Performance of Optimized U-Net Model}
\begin{tabular}{|c|c|c|c|}
\hline
Platform & Hardware & Inference Speed & Remarks \\
\hline
Apple Mac mini (M4 Pro) & CPU & $\sim$0.5 sec/frame & Not suitable for real-time use \\
\hline
Apple Mac mini (M4 Pro) & MPS (GPU) & $>$10 Hz & Efficient edge deployment \\
\hline
DESY Maxwell Cluster & GPU & Real-time (>$10$ Hz) & Suitable for online deployment \\

\hline
\end{tabular}
\label{tab:computational_performance}
\end{table}

\section{Conclusion}

This work presented a deep-learning-based framework for robust ROI detection in XFEL beam images using denoising autoencoders and U-Net segmentation models. Compared with conventional bounding-box and classical image-processing approaches, the proposed methods provide more accurate and reliable ROI localization for complex beam profiles, including low-intensity, distorted, and multiple-beam structures. While the autoencoder proved effective for beam denoising and feature enhancement, the optimized U-Net achieved superior pixel-level segmentation accuracy, and the multiclass U-Net successfully separated overlapping beam regions. The optimized U-Net satisfied the computational requirements for real-time accelerator applications and was successfully deployed within a prototype control-system server for continuous beam image analysis. Although trained exclusively on synthetic data, the models demonstrated robust performance on real XFEL beam images, highlighting the effectiveness of the synthetic data generation strategy. The proposed methodology provides a practical, scalable, and transferable solution for image-based beam diagnostics at the European XFEL and other accelerator facilities.

\section{Acknowledgments}

The work was undertaken as an Engineering Innovation Project at the Indian Institute of Technology Jodhpur, with a tie-up with the DESY Summer Student Programme 2025. Prachiti Sujit Chandratreya (PSC) and Jitendra Kumar thank the DESY team for providing beam image data and computing infrastructure support. This research was supported in part by the Maxwell computing resources at Deutsches Elektronen-Synchrotron DESY, Hamburg, Germany. PSC acknowledges the Office of Academics and the administration of the Indian Institute of Technology Jodhpur for supporting the Engineering Innovation Project, which enabled participation in the DESY Summer Student Programme 2025 and the associated research project. The authors acknowledge support from DESY (Hamburg, Germany), a member of the Helmholtz Association HGF and European XFEL GmbH (Schenefeld, Germany).

\bibliographystyle{JHEP}
\bibliography{biblio}

\end{document}